\documentclass[prl,aps,twocolumn,superscriptaddress,
]{revtex4}
\usepackage{amsmath,amssymb,graphicx}
\usepackage{pxfonts}
\usepackage{graphicx}
\usepackage{color}
\usepackage{float}
\begin{document}


\title{High-mobility spin transport mediated by indirect excitons in a van der Waals heterostructure}

\author{Zhiwen~Zhou$^*$}
\affiliation{Department of Physics, University of California San Diego, La Jolla, CA 92093, USA}
\author{W.~J.~Brunner$^*$}
\affiliation{Department of Physics, University of California San Diego, La Jolla, CA 92093, USA}
\author{Luc~M. Nguyen}
\affiliation{Department of Physics, University of California San Diego, La Jolla, CA 92093, USA}
\author{C.~A.~Lozano}
\affiliation{Department of Physics, University of California San Diego, La Jolla, CA 92093, USA}
\author{G.~Mercado}
\affiliation{Department of Physics, University of California San Diego, La Jolla, CA 92093, USA}
\author{L.~H.~Fowler-Gerace}
\affiliation{Department of Physics, University of California San Diego, La Jolla, CA 92093, USA}
\author{L.~V.~Butov} 
\affiliation{Department of Physics, University of California San Diego, La Jolla, CA 92093, USA}

\begin{abstract}
\noindent Efficient spin transport with high spin mobility is of both fundamental and practical significance for spintronics. Spin can be transferred by moving particles. However, particle scattering limits spin transfer and prevents high spin mobilities. 
In this work, we study the kinetics of spin transport mediated by a quantum system of indirect excitons (IXs) in a van der Waals heterostructure composed of atomically thin transition-metal dichalcogenides (TMDs). This spin transfer mechanism allows for the realization of efficient spin transport characterized by a high spin mobility reaching $10^5$~cm$^2$/(eV$\cdot$s), which is orders of magnitude higher than that achieved via standard diffusive spin transport in TMD materials. The suppression of IX scattering enables high-mobility spin transport. 
\end{abstract}
\maketitle

Spin transport is the foundation for a number of fundamental phenomena, including the spin Hall effect~\cite{Dyakonov1971, Hirsch1999, Sinova2004, Sih2005}, spin diffusion and drag~\cite{Kikkawa1999, DAmico2001, Weber2005, Crooker2005, Carter2006}, the persistent spin helix~\cite{Koralek2009}, and the quantum spin Hall effect~\cite{Kane2005, Bernevig2006, Konig2007}. In addition to its fundamental significance, efficient spin transport is crucial for spintronic devices, which may offer advantages in energy efficiency and operation speed over conventional charge-based devices~\cite{Awschalom2007}.

Spin can be transferred by moving particles. However, particle scattering, in addition to suppressing their transport, causes spin relaxation; this limits the spin transfer and prevents high spin mobilities~\cite{Dyakonov2008}. 

Particle scattering can be suppressed in a quantum system of bosons due to superfluidity. Spatially indirect excitons (IXs), also known as interlayer excitons, provide a quantum system of bosons. An IX is composed of an electron and a hole confined in separated layers in a heterostructure~\cite{Lozovik1976}. Due to the layer separation, the lifetimes of IXs exceed the lifetimes of spatially direct excitons (DXs), typically by orders of magnitude~\cite{Zrenner2992}. The long lifetimes allow IXs to cool down below the temperature of quantum degeneracy and form a superfluid condensate~\cite{High2012}. Furthermore, the reduced overlap of the electron and hole wave functions in IXs suppresses the spin relaxation caused by electron-hole exchange~\cite{Maialle1993}. With these properties, IXs provide a platform for achieving high-mobility spin transport.

IX-mediated spin transport was studied in GaAs heterostructures~\cite{Leonard2009, High2013, Violante2015, Finkelstein2017, Leonard2018}. However, the low temperatures of quantum degeneracy for IXs in GaAs heterostructures limit efficient IX-mediated spin transport in this material to a few Kelvin~\cite{High2013}. The temperature of quantum degeneracy, which can be achieved for excitons, scales proportionally to the exciton binding energy~\cite{Fogler2014}. High exciton binding energies can be achieved in van der Waals heterostructures composed of atomically thin layers of transition-metal dichalcogenides (TMDs)~\cite{Geim2013, Ye2014, Chernikov2014, Goryca2019}. IX binding energies in TMD heterostructures reach hundreds of meV~\cite{Fogler2014, Deilmann2018}, two orders of magnitude higher than in GaAs heterostructures~\cite{Sivalertporn2012}. Therefore, TMD heterostructures offer the opportunity to realize efficient IX-mediated spin transport at high temperatures. 

Due to the coupling of the spin and valley indices in TMD heterostructures~\cite{Xiao2012, Cao2012, Zeng2012, Mak2012}, spin transport is tied to valley transport; therefore, we also use the term spin transport to encompass spin-valley transport. Spin transport can be measured by detecting the transport of spin-polarized carriers via spatially and polarization-resolved photoluminescence (PL) imaging. This method was used to explore diffusive spin transport in systems of IXs~\cite{Rivera2016, Unuchek2019, Huang2020, Shanks2022, Zhou2024}, DXs~\cite{Onga2017}, and electrons or holes~\cite{Jin2018, Ren2022} in TMD heterostructures.

In this work, we study the kinetics of spin transport mediated by a quantum system of IXs in a van der Waals MoSe$_2$/WSe$_2$ heterostructure. We observe efficient spin transport characterized by a high spin mobility that reaches $10^5$~cm$^2$/(eV$\cdot$s).

\vskip5mm
{\bf Results}

IXs in the MoSe$_2$/WSe$_2$ heterostructure are composed of electrons and holes confined in adjacent MoSe$_2$ and WSe$_2$ monolayers~\cite{Rivera2015}. The heterostructure details are outlined in the Supplementary Information (SI). 

In the measurements, the focused, circularly polarized laser excitation generates spin-polarized IXs. The laser excitation energy is resonant to the DX in the WSe$_2$ heterostructure layer. The laser pulses have a step-like shape with an 89~ns pulse width, a 0.3~ns edge sharpness, and a 300~ns pulse period. The transport of spin-polarized IXs is measured by time-, polarization-, and spatially-resolved optical imaging of IX PL. The measurement details are outlined in the SI.

\begin{figure}
\begin{center}
\includegraphics[width=8.7cm]{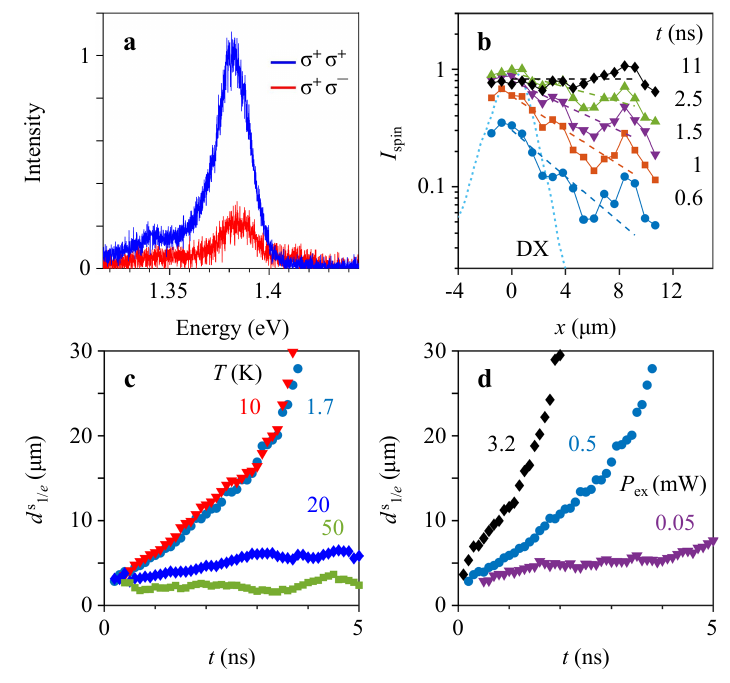}
\caption{Kinetics of IX-mediated spin transport. 
(a) The polarization of IX PL. The laser excitation is circularly polarized. The blue spectrum is co-polarized with the laser excitation ($\sigma^+ \sigma^+$), and the red spectrum is cross-polarized ($\sigma^+ \sigma^-$). 
(b) Spatial profiles of spin density ($I_{\rm spin} = I_{\sigma^+} - I_{\sigma^-}$) vs. time $t$ after the onset of the laser excitation pulse. The excitation spot is indicated by the cyan dotted line, which represents the DX PL profile in the MoSe$_2$ monolayer and closely matches the laser profile for short-range DX transport. 
The dashed lines show least-squares fits of the spin density profiles ($I_{\rm spin}(x)$) to exponential decays in the region from the excitation spot to the HS edge, $x = 0 - 9$~$\mu$m.
(c,d) The $1/e$ decay distance of the spin density profiles $d^{\rm s}_{1/e}$ vs. $t$ for different temperatures $T$ (c) and laser excitation powers $P_{\rm ex}$ (d). $P_{\rm ex} = 0.2$~mW, $T = 6$~K in (a). $P_{\rm ex} = 0.5$~mW, $T = 1.7$~K in (b). $P_{\rm ex} = 0.5$~mW in (c). $T = 1.7$~K in (d).
The efficient spin transport with rapid enhancement of $d^{\rm s}_{1/e}$ (c,d) is observed at temperatures and densities outlined in Fig.~3. 
}
\end{center}
\label{fig:spectra}
\end{figure}

\begin{figure}
\begin{center}
\includegraphics[width=8.7cm]{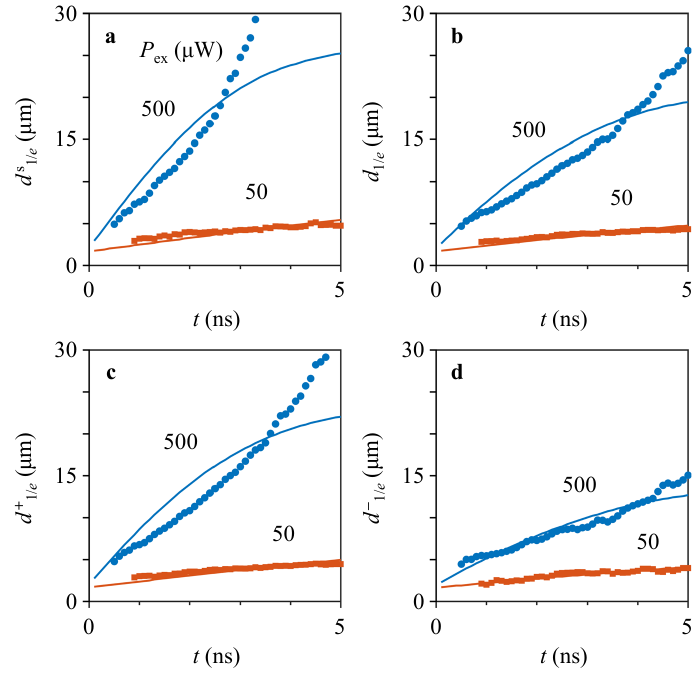}
\caption{Comparison of transport kinetics. 
(a) $d^{\rm s}_{1/e}(t)$: Spin transport kinetics.
(b) $d_{1/e}(t)$: IX transport kinetics.
(c) $d^+_{1/e}(t)$: Majority-spin IX transport kinetics.
(d) $d^-_{1/e}(t)$: Minority-spin IX transport kinetics.
$d^{\rm s}_{1/e}$, $d_{1/e}$, $d^+_{1/e}$, and $d^-_{1/e}$ are the $1/e$ decay distances of the spin density ($I_{\rm spin} = I_{\sigma^+} - I_{\sigma^-}$), IX density ($I = I_{\sigma^+} + I_{\sigma^-}$), majority-spin IX density ($I_{\sigma^+}$), and minority-spin IX density ($I_{\sigma^-}$), respectively. 
$\sigma^+$ is co-polarized and $\sigma^-$ is cross-polarized with the laser excitation. 
The data for excitation power $P_{\rm ex} = 500$~$\mu$W (50~$\mu$W) are shown by blue points (red squares). $T = 3.5$~K. The laser excitation pulse starts at $t = 0$. 
Lines represent the simulations. In the simulations, the mobilities $\mu_{\rm s}$ for spin transport, $\mu$ for IX transport, $\mu_+$ for majority-spin IX transport, and $\mu_-$ for minority-spin IX transport are estimated by a least-squares fit of the expansion kinetics simulated with Eq.~1 to the measured $d^{\rm s}_{1/e}(t)$, $d_{1/e}(t)$, $d^+_{1/e}(t)$, and $d^-_{1/e}(t)$, respectively. 
The obtained spin mobility $\mu_{\rm s}$ is shown in Fig.~3. $\mu$, $\mu_+$, and $\mu_-$ are shown in Fig.~S7 in the SI. 
}
\end{center}
\label{fig:spectra}
\end{figure}

The PL intensity of IXs co-polarized with the laser excitation ($I_{\sigma^+}$) significantly exceeds that of cross-polarized IXs ($I_{\sigma^-}$), indicating a strong spin polarization created by the laser excitation (Fig.~1a). Spin-polarized IXs propagating from the laser excitation spot transfer the spin polarization. The degree of circular polarization of IX PL is given by $P = (I_{\sigma^+} - I_{\sigma^-}) / (I_{\sigma^+} + I_{\sigma^-})$. The transport of IX density ($n$), referred to as IX transport, is characterized by the propagation of the quantity $I = I_{\sigma^+} + I_{\sigma^-}$. The transport of spin polarization density, referred to as spin transport, is characterized by the propagation of the quantity $I_{\rm spin} = Pn = I_{\sigma^+} - I_{\sigma^-}$.

Figure~1b shows the kinetics of spin density profiles ($I_{\rm spin}(x)$). Spin transport is characterized by the $1/e$ decay distance of the spin polarization density ($d^{\rm s}_{1/e}$), which is determined by a least-squares exponential fit to the $I_{\rm spin}(x)$ data, as shown in Fig.~1b. The evolution of $d^{\rm s}_{1/e}(t)$ quantifies the kinetics of spin transport. Figures 1c and 1d show $d^{\rm s}_{1/e}(t)$ for different temperatures $T$ and laser excitation powers $P_{\rm ex}$. A rapid enhancement of $d^{\rm s}_{1/e}(t)$, observed within a specific range of temperatures and densities, indicates efficient spin transport kinetics, as outlined below. 

For the characterization of spin transport kinetics, we use a drift-diffusion approach (other approaches considering diffusion or ballistic transport are outlined in the SI, and they give qualitatively similar results). This approach allows the estimation of the spin mobility ($\mu_{\rm s}$) and enables a comparison of spin transport across different parameters. We note that while this approach characterizes spin transport and estimates spin mobility, it is not intended to precisely describe the efficient spin transport outlined below.

\begin{figure}
\begin{center}
\includegraphics[width=8.7cm]{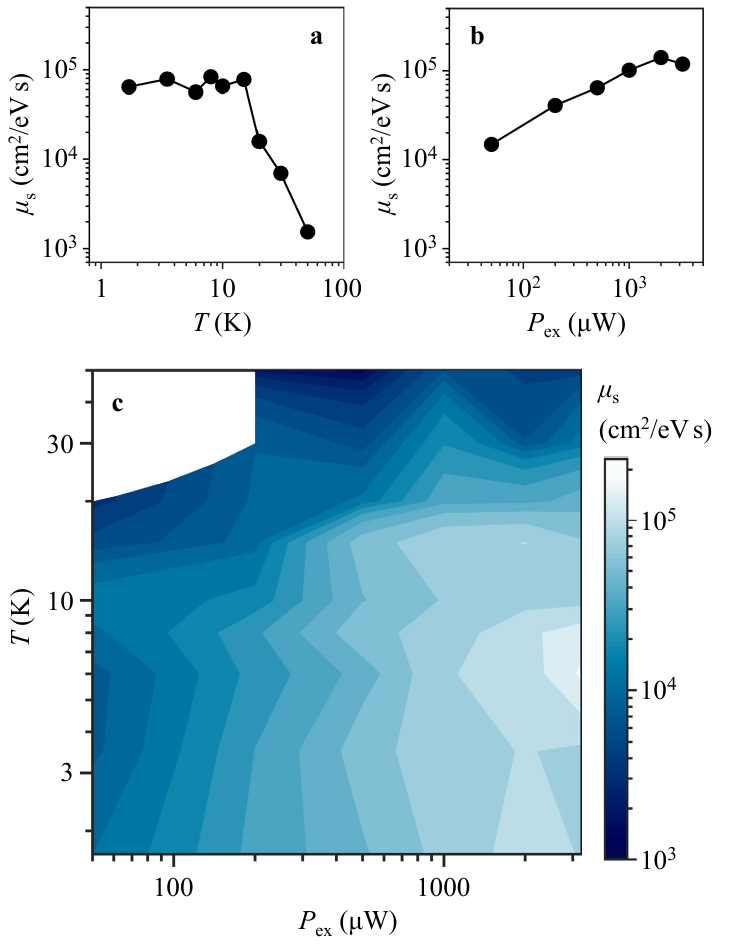}
\caption{Spin mobility. The spin mobility $\mu_{\rm s}$ is estimated by a least-squares fit of the expansion kinetics simulated with Eq.~1 to the measured $d^{\rm s}_{1/e}(t)$. (a-c): The estimated $\mu_{\rm s}$ vs. temperature $T$ (a), excitation power $P_{\rm ex}$ (b), and both $T$ and $P_{\rm ex}$ (c). 
$P_{\rm ex} = 0.5$~mW in (a). $T = 1.7$~K in (b).
}
\end{center}
\label{fig:spectra}
\end{figure}

The spin transport kinetics are simulated using the equation 
\begin{eqnarray}
\frac{\partial n_i}{\partial t} = \nabla \left[ D_i \nabla n_i + \mu_i n_i \nabla (n u_0 + U) \right] + \Lambda_i - \frac{n_i}{\tau}\ 
\end{eqnarray}
The first term describes diffusion, the second and third terms describe drift due to the IX interaction and the in-plane potential energy landscape, and the last two terms represent generation by laser excitation and recombination~\cite{Ivanov2002}. For the majority-spin and minority-spin IXs, $n_i$ represents their densities, which are given by $I_{\sigma^+}$ and $I_{\sigma^-}$, respectively. In turn, $n_i$ is given by $I_{\sigma^+} - I_{\sigma^-}$ for spin transport and by $I_{\sigma^+} + I_{\sigma^-}$ for IX transport. The spatial profile of the IX generation, $\Lambda_i$, corresponds to the laser excitation profile shown by the dotted line in Fig.~1b, its temporal profile matches that of the laser pulses, and the ratio between the generation of minority-spin and majority-spin IXs, $\Lambda_-/\Lambda_+$, matches the ratio between the cross-polarized and co-polarized IX PL intensities shown in Fig.~1a. The measurement of in-plane potential energy landscape $U$ caused by variations of the local IX environment in the heterostructure is presented in the SI. The measurement of IX lifetime $\tau$ is presented in the SI. We use the Einstein relation for diffusivity $D_i = \mu_i k_{\rm B}T$ ($k_{\rm B}$ is the Boltzmann constant). The IX interaction is approximated by the mean-field model~\cite{Yoshioka1990} using the measured IX energy shift $\delta E = nu_0$ with $u_0 = 4 \pi e^2 d_z/ \varepsilon$ as outlined in the SI ($d_z \sim 0.6$~nm is the separation between the electron and hole layers and $\varepsilon \sim 7.4$ is the dielectric constant for the MoSe$_2$/WSe$_2$ heterostructure~\cite{Laturia2018}). Mobility is the only fitting parameter in Eq.~1. 

Figure~2 shows the spin transport kinetics $d^{\rm s}_{1/e}(t)$ (Fig.~2a), the IX transport kinetics $d_{1/e}(t)$ (Fig.~2b), the majority-spin IX transport kinetics $d^+_{1/e}(t)$ (Fig.~2c), and the minority-spin IX transport kinetics $d^-_{1/e}(t)$ (Fig.~2d). $d^{\rm s}_{1/e}$, $d_{1/e}$, $d^+_{1/e}$, and $d^-_{1/e}$ are the $1/e$ decay distances of the spin density ($I_{\rm spin} = I_{\sigma^+} - I_{\sigma^-}$), IX density ($I = I_{\sigma^+} + I_{\sigma^-}$), majority-spin IX density ($I_{\sigma^+}$), and minority-spin IX density ($I_{\sigma^-}$), respectively. $\sigma^+$ is co-polarized and $\sigma^-$ is cross-polarized with the laser excitation. Lines represent the simulations. In the simulations, the mobilities $\mu_{\rm s}$ for spin transport, $\mu$ for IX transport, $\mu_+$ for majority-spin IX transport, and $\mu_-$ for minority-spin IX transport are estimated by a least-squares fit of the expansion kinetics simulated with Eq.~1 to the measured $d^{\rm s}_{1/e}(t)$, $d_{1/e}(t)$, $d^+_{1/e}(t)$, and $d^-_{1/e}(t)$, respectively. The obtained spin mobility $\mu_{\rm s}$ is shown in Fig.~3 ($\mu$, $\mu_+$, and $\mu_-$ are shown in Fig.~S7 in the SI). 

Figure~3 shows that the spin mobility reaches high values of $\sim 10^5$~cm$^2$/(eV$\cdot$s). The high spin mobility vanished at temperatures above $\sim 10$~K (Fig.~3a). The spin mobility first increases with density and then starts to decline at the highest excitation densities in the experiment (Fig.~3b). A qualitatively similar dependence on temperature and excitation density is observed for the IX mobility, as shown in Fig.~S7 in the SI.

\begin{figure}
\begin{center}
\includegraphics[width=8.7cm]{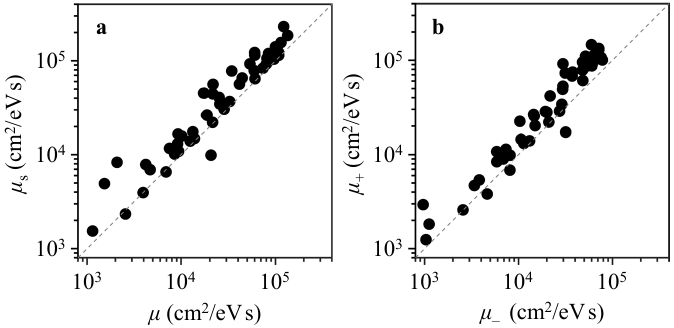}
\caption{Comparison of mobilities.
(a) Spin mobility vs. IX mobility, $\mu_{\rm s}$ vs. $\mu$. The values of $\mu_{\rm s}$ and $\mu$ are taken from Fig.~3c and Fig.~S7a, respectively. The enhancement of $\mu_{\rm s}$ with increasing $\mu$ is observed across a broad range of both excitation densities and temperatures corresponding to the parameter ranges in Fig.~3c. The dashed line indicates the diagonal $\mu_{\rm s} = \mu$. In the high-mobility regime, the mobility of IX-mediated spin transport exceeds the IX mobility, $\mu_{\rm s}> \mu$.  
(b) Mobility of majority-spin IXs vs. mobility of minority-spin IXs, $\mu_+$ vs. $\mu_-$. The values of $\mu_+$ and $\mu_-$ are taken from Fig.~S7b and Fig.~S7c, respectively. The dashed line indicates the diagonal $\mu_+ = \mu_-$. In the high-mobility regime, the mobility of majority-spin IXs exceeds the mobility of minority-spin IXs, $\mu_+ > \mu_-$.
}
\end{center}
\label{fig:spectra}
\end{figure}

A comparison of Figs.~2a and 2b shows that the spin mobility exceeds the IX mobility ($\mu_{\rm s} > \mu$), while a comparison of Figs.~2c and 2d reveals that the mobility of majority-spin IXs exceeds the mobility of minority-spin IXs ($\mu_+ > \mu_-$). These trends hold across a wide range of temperatures and excitation densities, as confirmed by the data in Figs.~3 and S7.

Figure~4a shows the correlation between the spin mobility $\mu_{\rm s}$ and IX mobility $\mu$. The enhancement of $\mu_{\rm s}$ with increasing $\mu$ is observed across a broad range of both excitation densities and temperatures corresponding to the parameter ranges in Fig.~3c. Figure~4a also shows that in the high-mobility regime, the spin mobility exceeds the IX mobility ($\mu_{\rm s} > \mu$), which is consistent with the comparisons between Figs.~2a and 2b, and Figs.~3c and S7a, as outlined above. Figure~4b shows that in the high-mobility regime, the mobility of majority-spin IXs exceeds the mobility of minority-spin IXs ($\mu_+ > \mu_-$), which is consistent with the comparisons between Figs.~2c and 2d, and Figs.~S7b and S7c, as outlined above.

\vskip5mm
{\bf Discussion}

Earlier studies with unpolarized optical excitation reported high IX mobility values in the presence of substantial in-plane disorder, consistent with theoretical predictions of IX superfluidity~\cite{Fisher1989, Capogrosso-Sansone2008}, as outlined in Ref.~\cite{Zhou2026}. The measurements in this work are performed using the same heterostructure along the same IX path. This allows for data comparison. The high IX mobility $\mu$ measured with circularly polarized excitation in this work (Fig.~S7a) is consistent with the earlier data with unpolarized excitation~\cite{Zhou2026}, as outlined in the SI.

For spin transport mediated by a quantum system of IXs (Figs.~1-4), the observed spin mobility $\mu_{\rm s}$, reaching $10^5$~cm$^2$/(eV$\cdot$s) (Fig.~3), is significantly higher than the values achieved with conventional diffusive spin transport. For comparison, the estimated spin mobility for diffusive spin transport in IX, hole, and electron systems in TMD heterostructures was $\sim 50, 160$, and $3 \cdot 10^3$~cm$^2$/(V$\cdot$s)~\cite{units}, respectively~\cite{Rivera2016, Jin2018, Ren2022}.

The spin mobility $\mu_{\rm s}$ correlates with the IX mobility $\mu$, as shown in Fig.~4a. The enhancement of $\mu_{\rm s}$ with increasing $\mu$ and, in turn, with increasing mean free time $\tau_{\rm m}$ ($\tau_{\rm m} \sim \mu m$ where $m$ is the IX mass) is consistent with the reduction of spin relaxation due to the suppression of scattering of spin-carrying particles~\cite{Dyakonov2008}, as outlined in the introduction. The suppression of IX scattering, characterized by enhanced IX mobility and mean free time, enables high-mobility spin transport. 

In the high-mobility regime, the spin mobility exceeds the IX mobility ($\mu_{\rm s} > \mu$), as shown in Figs.~2a,b and 4a. This is consistent with the mobility of majority-spin IXs exceeding the mobility of minority-spin IXs ($\mu_+ > \mu_-$), as shown in Figs.~2c,d and 4b. Indeed, the spin asymmetry in mobility, characterized by a slower growth of $d^-_{1/e}(t)$ for minority-spin IXs compared to $d^+_{1/e}(t)$ for majority-spin IXs (Figs.~2c,d), increases the polarization $P$ and, in turn, $d^{\rm s}_{1/e}(t)$, thus enhancing the spin mobility.

The spin asymmetry in mobility is qualitatively discussed below. The considered IX system is strongly spin-polarized and the density of majority-spin IXs significantly exceeds that of minority-spin IXs (Fig.~1a). The scattering of majority-spin IXs conserves their total momentum, causing no substantial suppression of majority-spin IX transport. A lower mobility of minority-spin IXs indicates that their scattering with majority-spin IXs suppresses their transport. 
We confirmed that changing the excitation polarization from $\sigma^+$ to $\sigma^-$ exchanges the majority- and minority-spin IXs and their transport, while maintaining the high spin mobility values.

In summary, we studied the kinetics of spin transport mediated by a quantum system of indirect excitons in a van der Waals MoSe$_2$/WSe$_2$ heterostructure. We observed efficient spin transport characterized by a spin mobility significantly higher than that of standard diffusive spin transport in TMD materials.

\vskip5mm
{\bf Acknowledgements.} 
We thank M.~M.~Fogler and D.~E.~Parker for discussions and A.~K.~Geim for teaching us manufacturing TMD heterostructures. The studies were supported by NSF Grant 2516006. The heterostructure manufacturing was supported by NSF Grant 1905478. The data analysis was supported by the Department of Energy, Office of Basic Energy Sciences, under award DE-FG02-07ER46449.

\vskip3mm
$^*$ These authors contributed equally: W.~J.~Brunner, Zhiwen~Zhou.


\begin{references}
\bibitem{units}
The mobility units used in this work, cm$^2$/(eV$\cdot$s), correspond to neutral particles, such as excitons. The mobility units for charged particles, cm$^2$/(V$\cdot$s), are sometimes applied to excitons, which typically implies a conversion to neutral-particle mobility via a factor of $e$ (the elementary charge).
\end{references}
\end{document}



\title{Supporting Information for\\
High-mobility spin transport mediated by indirect excitons in a van der Waals heterostructure}

\author{Zhiwen~Zhou$^*$}
\affiliation{Department of Physics, University of California San Diego, La Jolla, CA 92093, USA}
\author{W.~J.~Brunner$^*$}
\affiliation{Department of Physics, University of California San Diego, La Jolla, CA 92093, USA}
\author{Luc~M. Nguyen}
\affiliation{Department of Physics, University of California San Diego, La Jolla, CA 92093, USA}
\author{C.~A.~Lozano}
\affiliation{Department of Physics, University of California San Diego, La Jolla, CA 92093, USA}
\author{G.~Mercado}
\affiliation{Department of Physics, University of California San Diego, La Jolla, CA 92093, USA}
\author{L.~H.~Fowler-Gerace}
\affiliation{Department of Physics, University of California San Diego, La Jolla, CA 92093, USA}
\author{L.~V.~Butov} 
\affiliation{Department of Physics, University of California San Diego, La Jolla, CA 92093, USA}

\begin{abstract}
\end{abstract}
\maketitle

\renewcommand*{\thefigure}{S\arabic{figure}}
\renewcommand{\theequation}{S\arabic{equation}}

{\bf Heterostructure}

The MoSe$_2$/WSe$_2$ heterostructure (Fig.~S1) is fabricated using the dry-transfer peel technique~\cite{Withers2015}. The manufacturing details are described in Ref.~\cite{Fowler-Gerace2024}, where the same heterostructure was used for cw studies of IXs. The MoSe$_2$ monolayer is on top of the WSe$_2$ monolayer. The thickness of the bottom hBN layers is $\sim 40$~nm and that of the top layer is $\sim 30$~nm. The WSe$_2$ and MoSe$_2$ flakes are rotationally aligned using their long edges (Fig.~S1b). The twist angle between the WSe$_2$ and MoSe$_2$ monolayers ($\delta \theta = 1.1^\circ$), which corresponds to a moir{\'e} superlattice period of $b = 17$~nm, agrees with the angle between the MoSe$_2$ and WSe$_2$ edges, as outlined in Ref.~\cite{Fowler-Gerace2024}. The accuracy of $\delta \theta$ estimates using long WSe$_2$ and MoSe$_2$ edges is comparable to that using SHG. We do not use SHG for additional estimates of $\delta \theta$ since the intense excitation pulses in SHG measurements may cause deterioration of the heterostructure. Moir{\'e} potentials may be affected by atomic reconstruction~\cite{Weston2020, Rosenberger2020, Zhao2023} and disorder and vary over the heterostructure area. The measured IX g-factor $g \sim - 16$ corresponds to H stacking~\cite{Seyler2019, Wozniak2020}.

\begin{figure}
\begin{center}
\includegraphics[width=8.5cm]{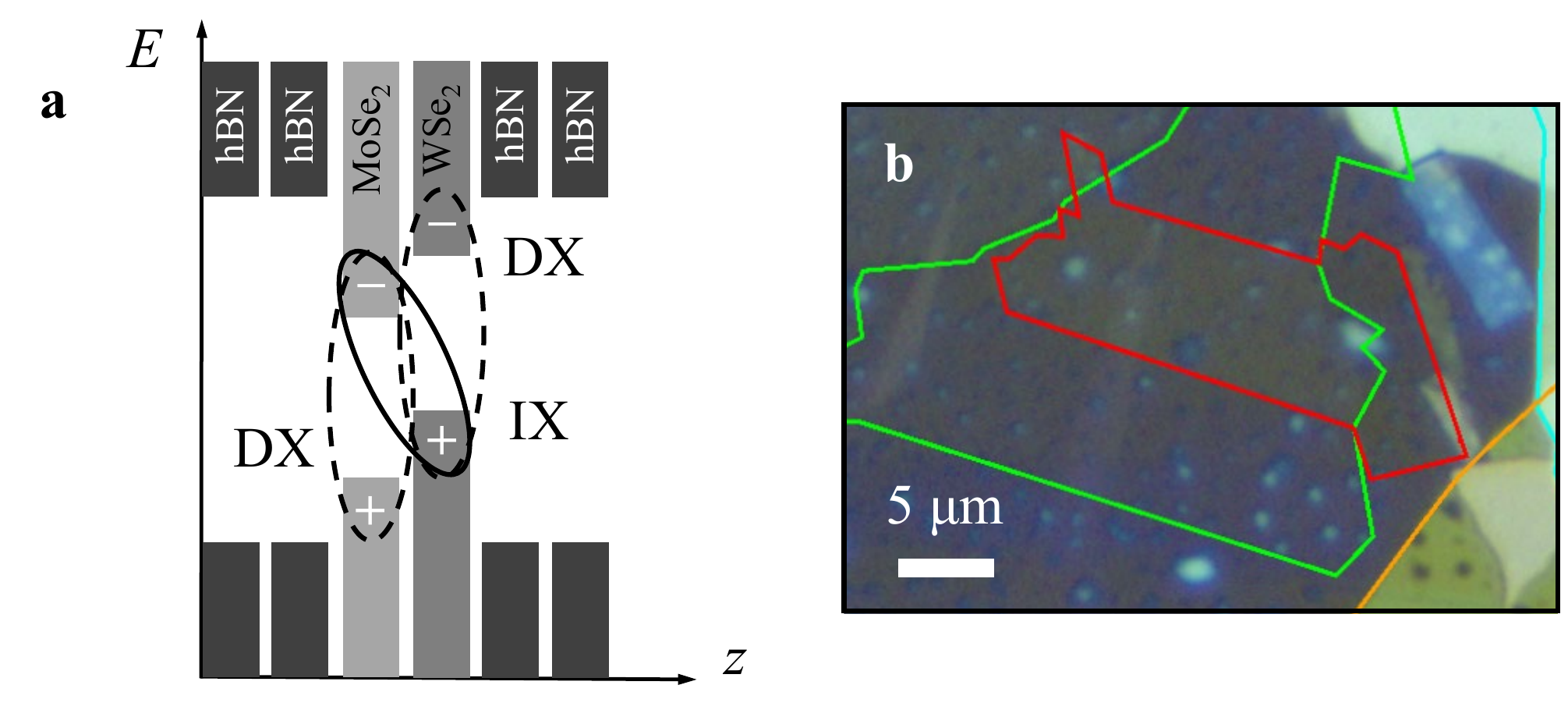}
\caption{(a) Energy-band diagram of the MoSe$_2$/WSe$_2$ van der Waals heterostructure.
The solid oval indicates IX, the dashed ovals DXs, (–) electrons, and (+) holes. (b) A microscope image of the heterostructure, with green, red, cyan, and orange lines indicating the boundaries of the WSe$_2$ monolayer, MoSe$_2$ monolayer, bottom hBN layer, and top hBN layer, respectively. 
}
\end{center}
\label{fig:spectra}
\end{figure}

\vskip5mm
{\bf Measurements}

Excitons are generated by focused circularly-polarized laser excitation with the spot half-width at half-maximum $\sim 1$~$\mu$m (the spot profile is shown in Fig.~1b). The  excitation energy $E_{\rm ex} = 1.689$~eV is resonant with the DXs in the WSe$_2$ heterostructure layer. The laser pulses are rectangular with an 89~ns pulse width, a 0.1~ns rise time, and a 0.3~ns fall time (the pulses rise and fall by a factor of $e$ within 0.1 and 0.3~ns, respectively). The pulse period is 300~ns. 

The transport of spin-polarized IXs is measured by time-, polarization-, and spatially-resolved optical imaging of circularly-polarized IX PL. An avalanche photodiode (APD) is moved across the optical image plane to measure the IX PL. An image magnification of $92 \times$ and an APD window diameter of 100~$\mu$m result in a spatial resolution of 1.1~$\mu$m for the imaging setup. The time resolution is 0.1~ns.

PL spectra are measured using a spectrometer with 0.2~meV resolution and a liquid-nitrogen-cooled CCD. The experiments are performed in a 4He cryostat. The sample is mounted on an Attocube $xyz$ piezo-translation stage that allows the sample position to be adjusted relative to a focusing lens inside the cryostat. All phenomena in this work are reproducible after multiple cooling cycles from room temperature to 2~K.

\vskip5mm
{\bf Simulation parameters}

The simulations using Eq.~1 utilize the following parameters: the lifetime $\tau$, the in-plane potential energy landscape $U$, and the interaction strength $u_0$. These parameters are determined as described below.

\vskip3mm
{\it Lifetime.}
The IX recombination lifetime ($\tau$) is obtained from the decay kinetics of the spatially integrated IX signal after the end of the excitation pulse. The spatial integration minimizes the effect of IX transport on the kinetics, realizing a regime where the decay kinetics is governed by the IX recombination. The decay kinetics of the co-polarized ($I^{\sigma ^+}_{\rm int}$) and cross-polarized ($I^{\sigma ^-}_{\rm int}$) signals are similar; this is observed across various temperatures and densities (Fig.~S2). Therefore, the decay kinetics of the combined signal in both polarizations ($I_{\rm int}$) is used~\cite{Zhou2026}. The variations of $\tau$ with excitation power and temperature are included in the simulations, using measured values of $\tau$ that correspond to the considered $T$ and $P_{\rm ex}$. This approach encompasses all recombination processes. The recombination rate, given by the slope of $\ln(I_{\rm int})$, decreases over time (Fig.~S2). The initial recombination rate after the end of the excitation pulse ($\tau^{-1}$) corresponds to the IX density during the pulse. Because the experiments measure IX transport during the pulse, the simulations use the initial lifetime $\tau$.

\begin{figure}
\begin{center}
\includegraphics[width=7.5cm]{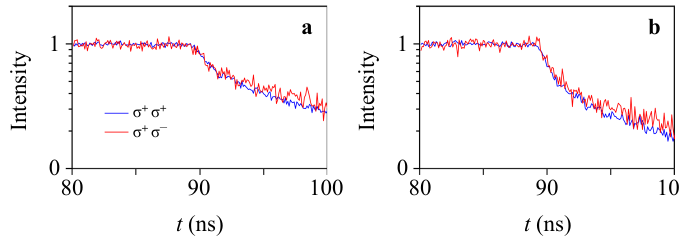}
\caption{Lifetime. The decay kinetics of the spatially integrated IX signal after the end of the excitation pulse, co-polarized (blue) and cross-polarized (red) with the laser excitation, for different excitation powers ($P_{\rm ex}$) and temperatures ($T$). $P_{\rm ex} = 0.2$~mW, $T = 1.7$~K in (a). $P_{\rm ex} = 0.5$~mW, $T = 6$~K in (b). The excitation pulse ends at $t \sim 89$~ns. The decay kinetics are used to estimate $\tau$ as outlined in the text.
}
\end{center}
\label{fig:spectra}
\end{figure}

\vskip3mm
{\it In-plane potential energy landscape.}
Equation~1 includes a term describing IX drift due to the in-plane potential energy landscape in the heterostructure ($U$). This landscape originates from variations in the local IX environment in the heterostructure, and its specific origin is not essential for modeling IX transport using Eq.~1. The measured $U(x)$ is shown in Fig.~S3. $U(x)$ is similar for both the co-polarized and cross-polarized signals, decreasing by $\sim 10$~meV over a distance of $\sim 10$~$\mu$m along the IX path from the excitation spot to the heterostructure edge (Fig.~S3c). For the simulations using Eq.~1, $U(x)$ is approximated by an energy gradient of 1~meV/$\mu$m.

\begin{figure}
\begin{center}
\includegraphics[width=8.7cm]{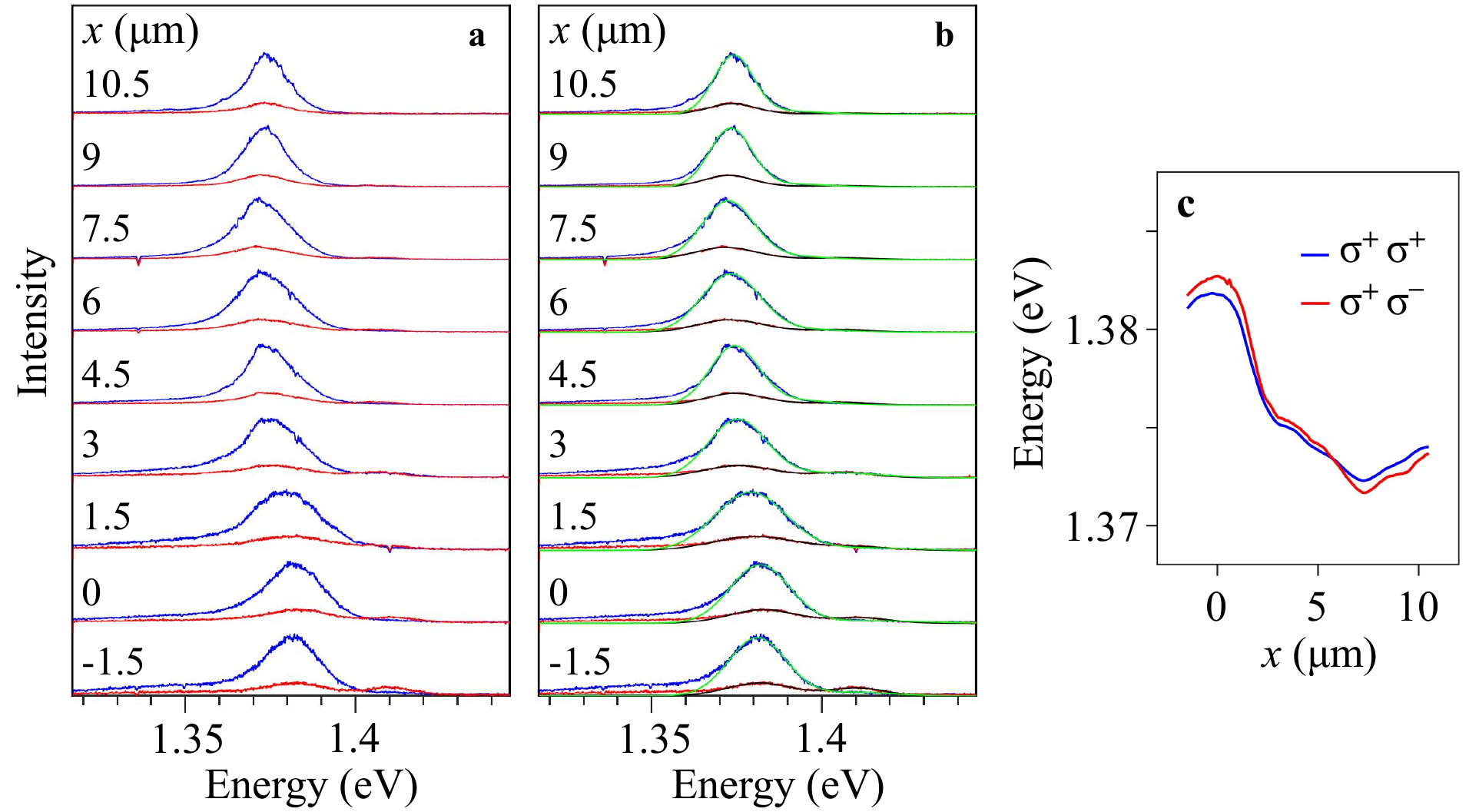}
\caption{Potential energy landscape. (a) The IX PL spectra, co-polarized (blue) and cross-polarized (red) with the laser excitation, vs. position $x$ along the IX path. 
(b) The same spectra with Gaussian fits, shown by the green lines for co-polarized spectra and by the black lines for cross-polarized spectra. 
(c) Position dependence of the IX energy $U(x)$ for the co-polarized (blue) and cross-polarized (red) PL spectra. 
The laser excitation is defocused to a $\sim 25$~$\mu$m spot covering the entire heterostructure. The power of this defocused excitation is $\sim 5$~mW. $T = 4.2$~K.
}
\end{center}
\label{fig:spectra}
\end{figure}

\begin{figure}
\begin{center}
\includegraphics[width=8.7cm]{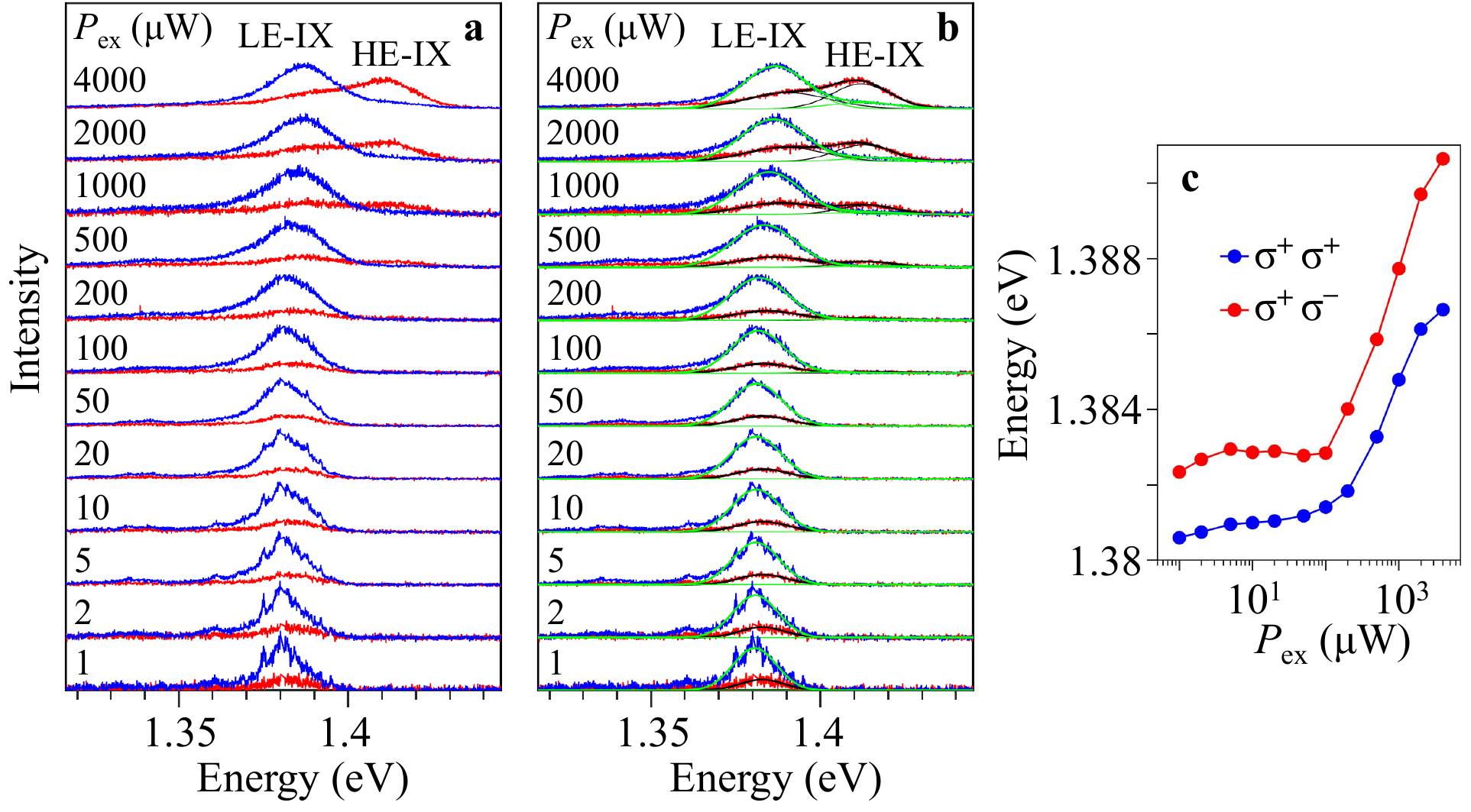}
\caption{IX spectra. (a) Excitation power dependence of the co-polarized (blue) and cross-polarized (red) IX spectra. The $\sim 2$~$\mu$m laser spot is centered at $x = 0$. $T = 6$~K.
The low-energy (LE) IX PL is co-polarized with the laser excitation and the high-energy (HE) IX PL is cross-polarized. The HE-IXs appear in the spectra at high $P_{\rm ex}$. The intensities are normalized.
(b) The same spectra with Gaussian fits, shown by the thin green lines for co-polarized spectra and by the thin black lines for cross-polarized spectra. The sum of the Gaussians is shown by the thick green and black lines, respectively. 
(c) Dependence of the co-polarized (blue) and cross-polarized (red) LE-IX energy on $P_{\rm ex}$.
}
\end{center}
\label{fig:spectra}
\end{figure}

\vskip3mm
{\it Interaction.}
IXs are characterized by out-of-plane electric dipoles. As shown in Ref.~\cite{Yoshioka1990}, the repulsive interaction between IXs can be approximated by the mean-field formula for the IX energy shift with density, $\delta E = nu_0 = n \frac{4 \pi e^2 d_z}{\varepsilon}$, where $d_z$ is the separation between the electron and hole layers and $\varepsilon$ is the dielectric constant. The mean-field model~\cite{Yoshioka1990} is commonly used to estimate IX densities in both GaAs and TMD heterostructures (e.g., in Refs.~\cite{Butov1999, Larionov2000, Ivanov2002, Butov2002, Vogele2009, Alloing2012, Jauregui2019, Unuchek2019, Sun2022, Zhang2024, Fowler-Gerace2024}), and is also employed for density estimation in this work. Including correlations lowers $u_0$ compared to the mean-field value, which increases the estimated $n$ for a given $\delta E$~\cite{Schindler2008, Remeika2009, Laikhtman2009, Ivanov2010, Remeika2015, Steinhoff2024}. 

The IX spectra showing the IX energy shift $\delta E$ are presented in Fig.~S4. A higher-energy IX (HE-IX) line appearing at high IX densities (Fig.~S4) was attributed to the appearance of moir{\'e} cells with double occupancy~\cite{Zhou2024}. IX transport is suppressed when the moir{\'e} cells occupancy $N$ reaches 1~\cite{Fowler-Gerace2024, Zhou2024, Zhou2026}. This is consistent with the Bose-Hubbard model, which predicts an insulating state, such as a Mott insulator or a Bose glass, when $N \sim 1$~\cite{Fisher1989}.

With increasing density, the HE-IX line appears in the heterostructure areas where $N$ reaches 1. Because transport is suppressed at $N \sim 1$, these areas are characterized by suppressed IX transport. This is confirmed by the lack of HE-IX transport: Fig.~S5 shows that the $1/e$ decay distances of HE-IX PL characterizing HE-IX transport (HE-$d_{1/e}$) and spin transport mediated by HE-IXs (HE-$d^{\rm s}_{1/e}$) are close to the spatial resolution. 

The shift $\delta E$ of the LE-IX line with $P_{\rm ex}$ is shown in Fig.~S4c. It is similar for both the co-polarized and cross-polarized signals. This shift $\delta E$ is used for the mean-field estimate of IX density $\delta E = nu_0$ following Ref.~\cite{Yoshioka1990} as outlined above. 

The heterostructure is elongated along the $x$ direction (Fig.~S1b). The IX transport is measured along $x$ and is approximated in the simulations as one-dimensional. Examples of simulated spatial profiles $I_{\rm spin}(x, t)$ for spin transport using Eq.~1 are shown in Fig.~S6. The simulations use the lifetime, the in-plane potential energy landscape, and the interaction outlined above.

\begin{figure}
\begin{center}
\includegraphics[width=8.7cm]{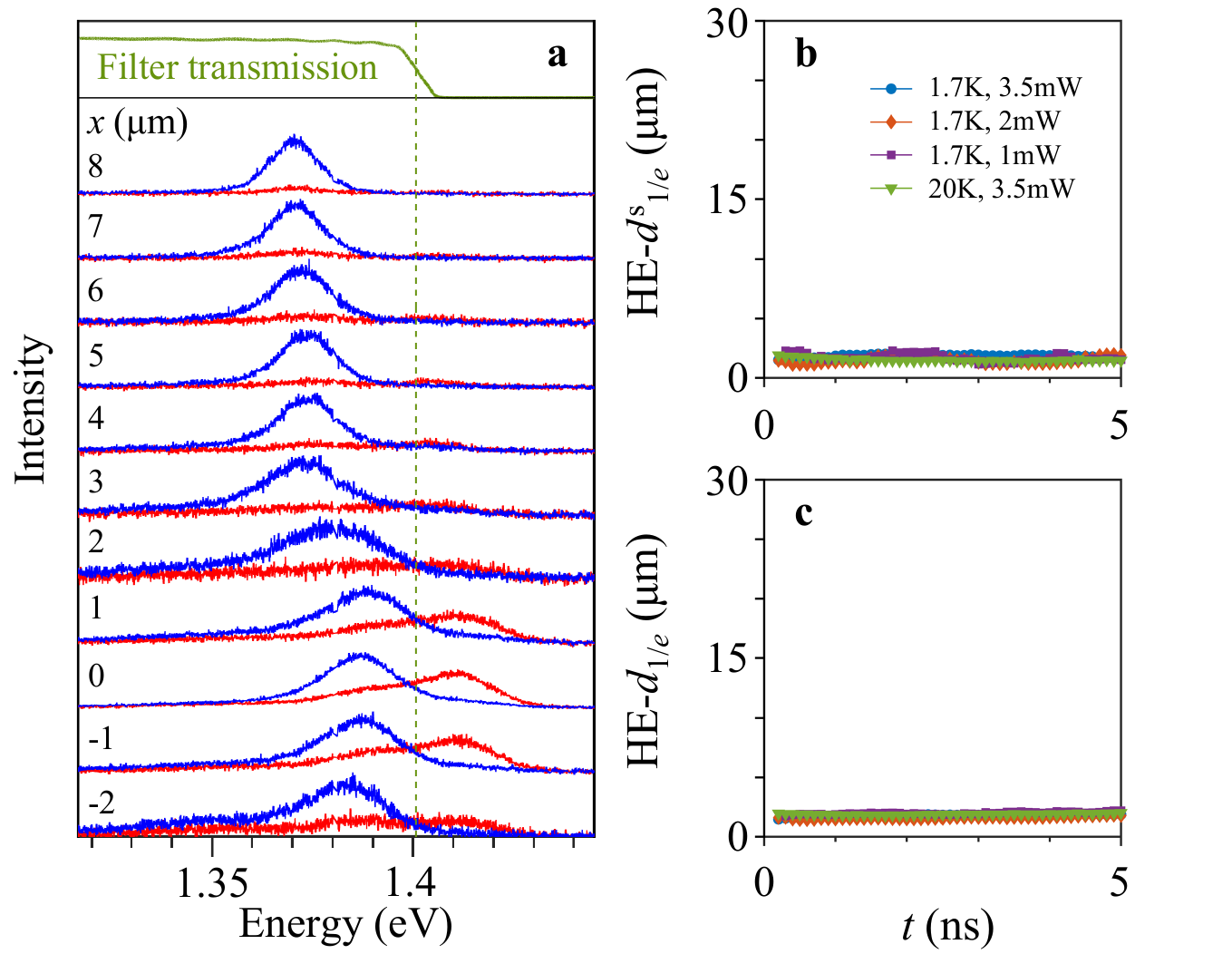}
\caption{HE-IXs. (a) The co-polarized (blue) and cross-polarized (red) IX PL spectra vs. position $x$. The $\sim 2$~$\mu$m laser spot is centered at $x = 0$. $P_{\rm ex} = 4$~mW. $T = 6$~K.
(b, c) The $1/e$ decay distances of HE-IX PL characterizing spin transport mediated by HE-IXs (HE-$d^{\rm s}_{1/e}$, (b)) and HE-IX transport (HE-$d_{1/e}$, (c)). 
The transmission spectrum of the filter used to select LE-IX PL in the transport measurements presented in Figs.~1-4 and S7-S10 is shown in (a).
}
\end{center}
\label{fig:spectra}
\end{figure}

\vskip5mm
{\bf Excitonic regime}

Figure~S4 shows that for all studied densities the LE-IX linewidth is close to that at the lowest densities. This indicates the excitonic regime for the studied transport kinetics. The excitonic regime is consistent with the IX density $n \sim 4 \times 10^{11}$~cm$^{-2}$ for the highest $P_{\rm ex}$ in the experiment, as estimated from the energy shift in Fig.~S4c. This density is significantly smaller than the density of the Mott transition to electron-hole plasma $n_{\rm Mott} > 10^{12}$~cm$^{-2}$~\cite{Fogler2014, Wang2019}. In contrast, the linewidth substantially increases in the plasma regime~\cite{Wang2019}. 

Narrow lines with the linewidth $\lesssim 1$~meV appear in the LE-IX spectrum at lower densities (Fig.~S4). These lines were attributed to localized states~\cite{Zhou2026a}.

\begin{figure}
\begin{center}
\includegraphics[width=5cm]{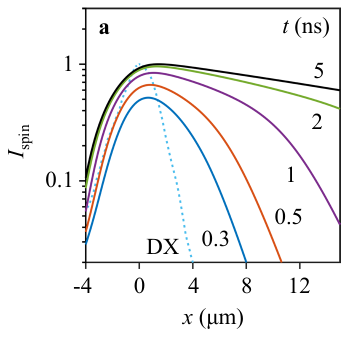}
\caption{Simulated spatial profiles $I_{\rm spin}(x, t)$ using Eq. 1. 
For these simulations, the spin mobility $\mu_{\rm s} = 7.9 \times 10^4$~cm$^2$/(eV$\cdot$s) was estimated by fitting the measured spin kinetics $d^{\rm s}_{1/e}(t)$ (blue points in Fig.~2a), as outlined in the main text. The dotted lines show the spatial profile of the laser excitation spot. The excitation pulse starts at $t = 0$.
}
\end{center}
\label{fig:spectra}
\end{figure}

\vskip5mm
{\bf IX, majority-spin IX, and minority-spin IX transport}

Figure~S7 shows the mobilities for IX transport ($\mu$), majority-spin IX transport ($\mu_+$), and minority-spin IX transport ($\mu_-$), estimated by a least-squares fit of the expansion kinetics simulated with Eq.~1 to the measured $d_{1/e}(t)$, $d^+_{1/e}(t)$, and $d^-_{1/e}(t)$, respectively. A comparison of the spin mobility $\mu_{\rm s}$ (Fig.~3) and IX mobility $\mu$ (Fig.~S7a) is shown in Fig.~4a and discussed in the main text. A comparison of the majority-spin IX mobility $\mu_+$ (Fig.~S7b) and minority-spin IX mobility $\mu_-$ (Fig.~S7c) is shown in Fig.~4b and discussed in the main text.

The high IX mobility $\mu$ measured with circularly polarized excitation in this work (Fig.~S7a) is consistent with the earlier data with unpolarized excitation in Ref.~\cite{Zhou2026}. In this work, the IX mobility is estimated using the $1/e$ decay distance of the LE-IX signal. Figure~S5a shows the transmission spectrum of the filter used to select LE-IX PL in the transport measurements. In Ref.~\cite{Zhou2026}, the IX mobility is estimated using the $1/e$ decay distance of the entire IX PL signal. Due to the lack of HE-IX transport (Fig.~S5c), the IX mobility estimated from the $1/e$ decay distance of the LE-IX signal is slightly higher than that estimated from the $1/e$ decay distance of the entire PL signal, which includes both LE-IXs and HE-IXs. We also note that the range of excitation powers in this work (up to $\sim 3$~mW) is smaller than that in Ref.~\cite{Zhou2026} (up to $\sim 4$~mW); the polarization optics reduces the maximum excitation power on the sample.

\begin{figure*}
\begin{center}
\includegraphics[width=17cm]{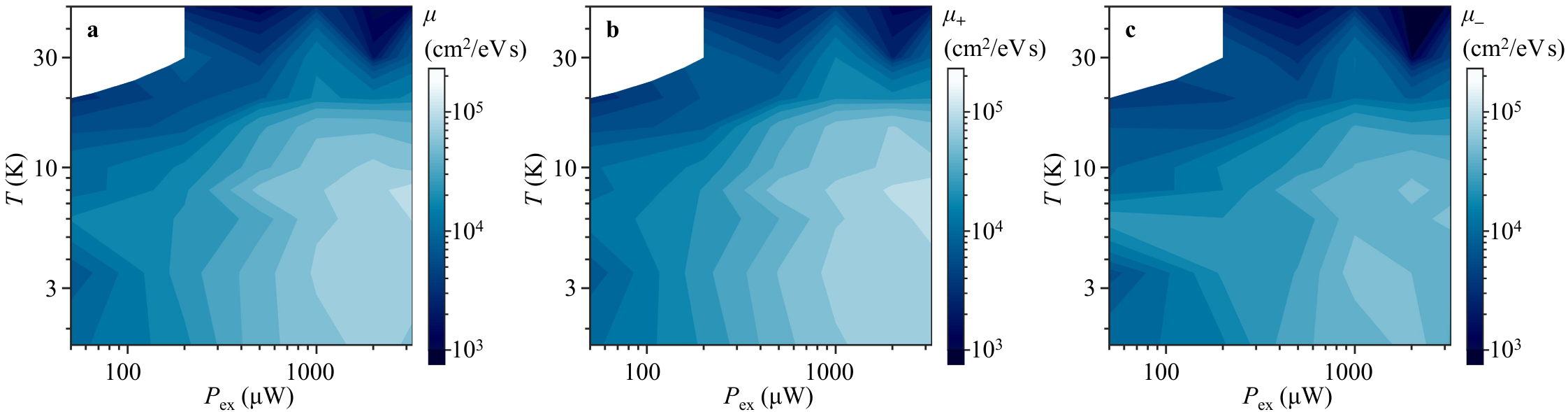}
\caption{IX, majority-spin IX, and minority-spin IX mobility. 
The mobilities $\mu$ for IX transport, $\mu_+$ for majority-spin IX transport, and $\mu_-$ for minority-spin IX transport are
estimated by a least-squares fit of the expansion kinetics simulated with Eq.~1 to the measured $d_{1/e}(t)$, $d^+_{1/e}(t)$, and $d^-_{1/e}(t)$, respectively.
(a-c): The estimated $\mu$ (a), $\mu_+$ (b), and $\mu_-$ (c) vs. temperature $T$ and excitation power $P_{\rm ex}$.
}
\end{center}
\label{fig:spectra}
\end{figure*}

\vskip5mm
{\bf Simulation fitting across different time ranges}

In the main text, the spin mobility is estimated by least-squares fit of the spin transport kinetics $d^{\rm s}_{1/e}(t)$ simulated with Eq.~1 to the measured $d^{\rm s}_{1/e}(t)$ in the time range $t = 0 - 5$~ns (Fig.~2a). 
Figure~S8a shows fits similar to those in Fig.~2a, but for the fit range $t = 0 - 3$~ns. These fits give the values of $\mu_{\rm s}$ shown in Figs.~S8b-d. The fits over $t = 0 - 3$~ns yield $\mu_{\rm s}$ values similar to those obtained from the fits over $t = 0 - 5$~ns (see Figs.~S8b-d and Fig.~3).

\begin{figure*}
\begin{center}
\includegraphics[width=17.9cm]{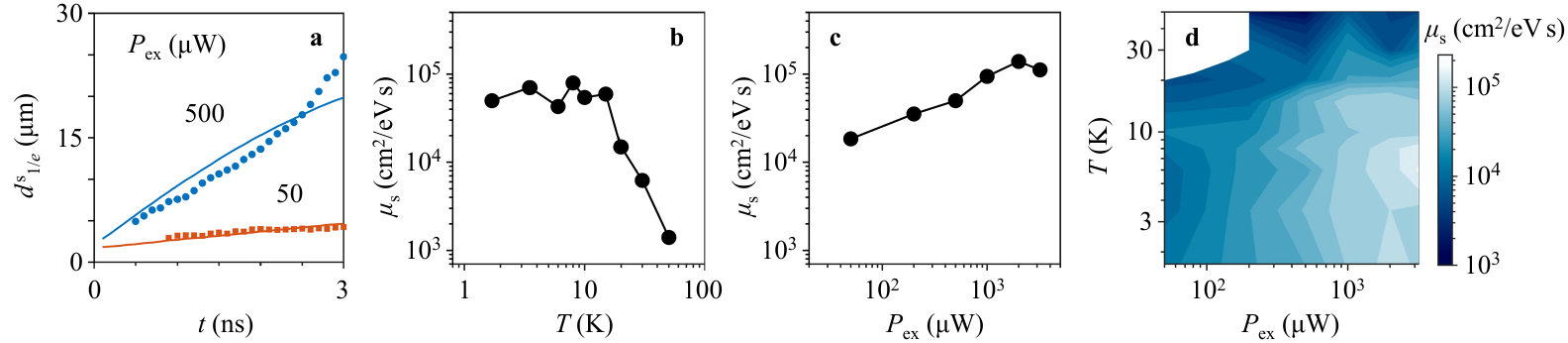}
\caption{Simulation using a $0 - 3$~ns fit range. (a) The measured spin transport kinetics $d^{\rm s}_{1/e}(t)$ is shown by blue points for the excitation power $P_{\rm ex} = 500$~$\mu$W and by red squares for $P_{\rm ex} = 50$~$\mu$W. $T = 3.5$~K. The simulated $d^{\rm s}_{1/e}(t)$ are shown by lines. 
(b-d) The estimated spin mobility $\mu_{\rm s}$ vs.  $T$ (b), $P_{\rm ex}$ (c), and both $T$ and $P_{\rm ex}$ (d). $P_{\rm ex} = 0.5$~mW in (b). $T = 1.7$~K in (c).
$\mu_{\rm s}$ is estimated by a least-squares fit of $d^{\rm s}_{1/e}(t)$ simulated with Eq.~1 to the measured $d^{\rm s}_{1/e}(t)$. 
The simulations in Fig.~S8 are similar to the simulations in Figs.~2a and 3, however, the fit range is $t = 0 - 3$~ns in Fig.~S8 and $t = 0 - 5$~ns in Figs.~2a and 3.
}
\end{center}
\label{fig:spectra}
\end{figure*}

\vskip5mm
{\bf Other approaches}

In the main text, the transport kinetics are simulated using the drift-diffusion approach. In this section, we consider simplified approaches: first, in terms of diffusion, and second, in terms of ballistic transport. The advantage of these two approaches lies in their simplicity; however, they neglect IX drift.

Within the diffusion approach, we characterize the measured spin transport kinetics using the diffusion equation $\frac{\partial n_{\rm s}}{\partial t} = D'_{\rm s} \Delta n_{\rm s} + \Lambda - \frac{n_{\rm s}}{\tau}$, where $D'_{\rm s}$ is the spin diffusivity and the other symbols correspond to those in Eq.~1. 
($D'_{\rm s}$ is used to avoid confusion with $D_{\rm s}$ in Eq. 1.)
$D'_{\rm s}$ is estimated by least-squares fit of the spin transport kinetics $d^{\rm s}_{1/e}(t)$ simulated with this equation to the measured $d^{\rm s}_{1/e}(t)$. The simulated $d^{\rm s}_{1/e}(t)$ are shows in Fig.~S9a and the obtained $D'_{\rm s}$ in Figs.~S9b-d. 

Within the ballistic transport approach, we characterize the measured spin transport kinetics using the ballistic transport equation $\frac{\partial n_{\rm s}}{\partial t} = - v_{\rm s} \nabla n_{\rm s} + \Lambda - \frac{n_{\rm s}}{\tau}$, where $v_{\rm s}$ is the spin transport velocity and the other symbols correspond to those in Eq.~1. $v_{\rm s}$ is estimated by least-squares fit of the spin transport kinetics $d^{\rm s}_{1/e}(t)$ simulated with this equation to the measured $d^{\rm s}_{1/e}(t)$. The simulated $d^{\rm s}_{1/e}(t)$ are shows in Fig.~S9e and the obtained $v_{\rm s}$ in Figs.~S9f-h. 

Simulations using the diffusion equation show nearly square-root-power growth of $d^{\rm s}_{1/e}(t)$ (Fig.~S9a). In contrast, those for drift-diffusion transport (Fig.~2a) and ballistic transport (Fig.~S9e) show nearly linear growth of $d^{\rm s}_{1/e}(t)$, consistent with the experiment. A nearly linear growth of $d^{\rm s}_{1/e}(t)$ indicates spin transport with nearly constant velocity, which is characteristic of both drift and ballistic transport. There are deviations from the square-root growth of $d^{\rm s}_{1/e}(t)$ for diffusion, and from the linear growth of $d^{\rm s}_{1/e}(t)$ for both drift-diffusion and ballistic transport. In particular, exciton recombination slows down this expansion.

All approaches show a similar dependence of spin transport kinetics on parameters. Efficient spin transport kinetics, characterized by high $\mu_{\rm s}$ (Fig.~3), $D'_{\rm s}$ (Fig.~S9b-d), and $v_{\rm s}$ (Fig.~S9f-h), are observed in the same range of temperatures and excitation densities.

\begin{figure*}
\begin{center}
\includegraphics[width=17.9cm]{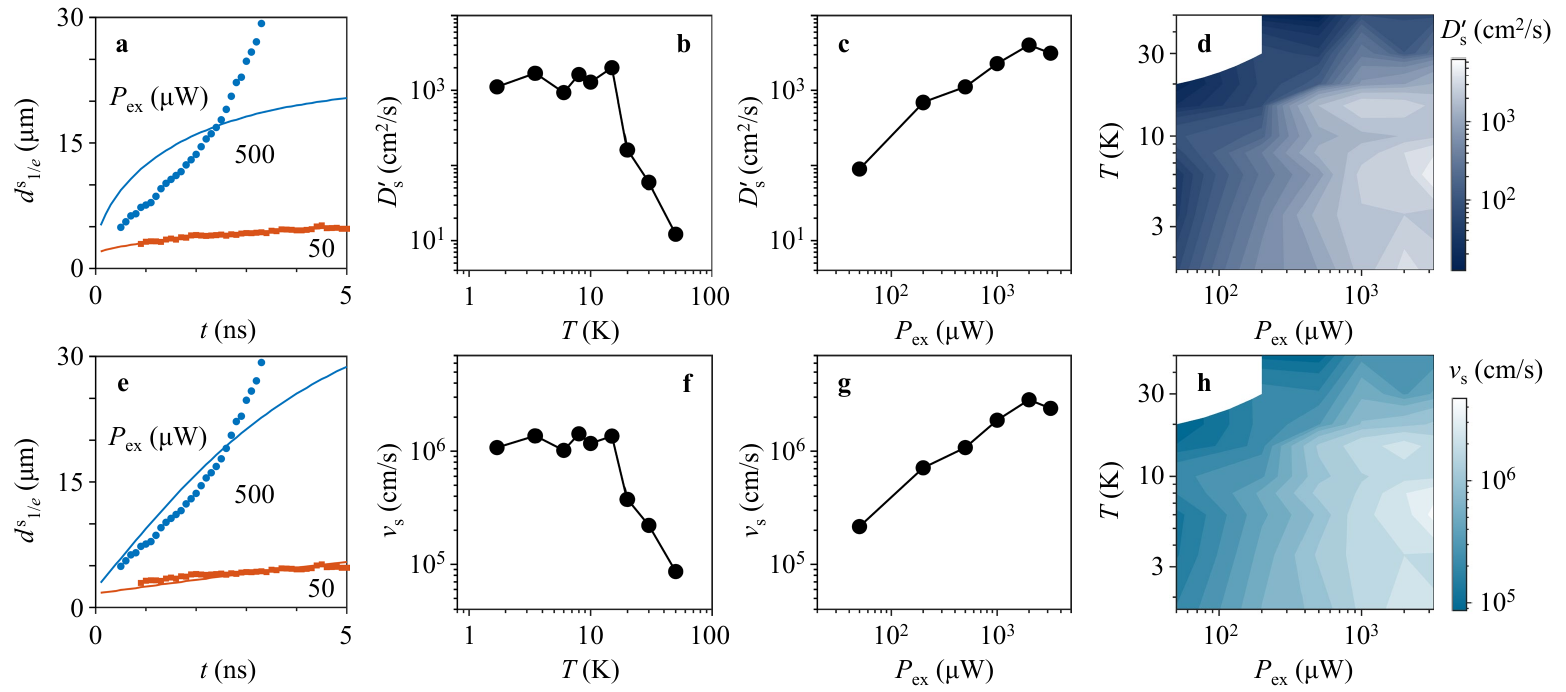}
\caption{Simulations using diffusion and ballistic transport approaches. (a,e) The measured spin transport kinetics $d^{\rm s}_{1/e}(t)$ for excitation power $P_{\rm ex} = 500$~$\mu$W (50~$\mu$W) are shown by blue points (red squares). $T = 3.5$~K. The simulated spin transport kinetics $d^{\rm s}_{1/e}(t)$ within the diffusion (a) and ballistic transport (e) approaches are shown by lines. 
(b-d) The spin diffusivity $D'_{\rm s}$ estimated by a least-squares fit of the spin transport kinetics $d^{\rm s}_{1/e}(t)$ simulated with the diffusion equation to the measured $d^{\rm s}_{1/e}(t)$.
(e-h) The spin velocity $v_{\rm s}$ estimated by a least-squares fit of the spin transport kinetics $d^{\rm s}_{1/e}(t)$ simulated with the ballistic transport equation to the measured $d^{\rm s}_{1/e}(t)$.
}
\end{center}
\label{fig:spectra}
\end{figure*}

\vskip10mm
{\bf Spin relaxation}

Spin transport in the main text and in Figs.~S7 and S8 is described by Eq.~1. This equation uses an approximation, which neglects spin relaxation. In this section, we consider the spin relaxation kinetics and its effect on IX-mediated spin transport. 

Spin relaxation processes typically suppress IX-mediated spin transport in comparison to IX transport~\cite{Leonard2009}. However, the data show efficient IX-mediated spin transport with the spin mobility exceeding the IX mobility (Figs.~2a, 2b, 3, and S7a). This indicates that the effect of spin relaxation on the suppression of the polarization transport is negligible during IX transport across the entire heterostructure ($\sim 10$~$\mu$m). 

The effect of spin kinetics on IX-mediated transport of spin polarization is outlined below. A similar issue was addressed in studies on GaAs heterostructures~\cite{Leonard2009}. Figure~S10a shows a schematic of the spin relaxation processes in GaAs heterostructures~\cite{Maialle1993}, where the heavy-hole exciton spin states include optically active (bright) states with $S_z = \pm 1$ and optically inactive (dark) states with $S_z = \pm 2$ composed of electron spin states $s_z = \pm 1/2$ and hole spin states $m_{\rm h} = \pm 3/2$.

Figure~S10b shows a schematic of spin relaxation processes in H-stacked MoSe$_2$/WSe$_2$ heterostructures based on the diagram of spin states in Refs.~\cite{Zhou2017, Wozniak2020}. The similarity between the spin relaxation processes in GaAs heterostructures (Fig.~S10a) and H-stacked MoSe$_2$/WSe$_2$ heterostructures (Fig.~S10b) justifies adapting the framework of spin transport with relaxation developed for GaAs heterostructures~\cite{Leonard2009} to the MoSe$_2$/WSe$_2$ heterostructures. 

The bright $e_{\downarrow - K}, h_{\uparrow + K}$ ($e_{\uparrow + K}, h_{\downarrow - K}$) excitons, dubbed b1 (b2) for brevity, are created by $\sigma^+$ ($\sigma^-$) polarized light and emit $\sigma^+$ ($\sigma^-$) polarized light. The $e_{\downarrow - K}, h_{\downarrow - K}$ and $e_{\uparrow + K}, h_{\uparrow + K}$ excitons, dubbed d1 and d2, are dark. The exciton PL polarization, $P =  (I_{\sigma^+} - I_{\sigma^-})/(I_{\sigma^+} + I_{\sigma^-})$, is determined by recombination and spin relaxation processes. An electron or hole spin-flip (and, in turn, a transfer from $K$ to $- K$ valley or vice versa) transforms a bright exciton into a dark state. The PL polarization decays when both the electron and hole flip their spins. This can occur in the two-step process due to the separate electron and hole spin flips, and in the single-step process due to the exciton spin flip (Fig.~S10b). 

In the absence of IX transport, the spin-flip model~\cite{Maialle1993}, adapted to the spin states in the MoSe$_2$/WSe$_2$ heterostructures (Fig.~S10b), gives
\begin{eqnarray}
\frac{\partial {\bar n}}{\partial t} = W {\bar n} + \bar{\Lambda}
\end{eqnarray}
where
\begin{widetext}
\begin{equation}
W = 
\begin{pmatrix}
   - (W^+_{\rm e} + W^+_{\rm h}) && W^-_{\rm h} && W^-_{\rm e} && 0 \\
    W^+_{\rm h} && -\left( \frac{1}{\tau} + W_{\rm ex} + W^-_{\rm e} + W^-_{\rm h} \right) && W_{\rm ex} && W^+_{\rm e} \\
    W^+_{\rm e} && W_{\rm ex} && -\left( \frac{1}{\tau} + W_{\rm ex} + W^-_{\rm e} + W^-_{\rm h} \right) && W^+_{\rm h} \\
    0 && W^-_{\rm e} && W^-_{\rm h} && - (W^+_{\rm e} + W^+_{\rm h})
\end{pmatrix}
,
\end{equation}
\end{widetext}
${\bar n} = \left( n_{\rm d1} \, \, n_{\rm b1} \, \, n_{\rm b2} \, \, n_{\rm d2} \right)^T$,  
$\bar{\Lambda} = \left( 0 \, \, \Lambda^+ \, \, \Lambda^- \, \, 0  \right)^T$,
$W_{\rm ex} = 1/(2 \tau_{\rm ex})$, 
$W^{\pm}_{\rm e(h)} = \tau_{\rm e(h)}^{-1} \left( 1 + e^{\pm \Delta / k_{\rm B} T} \right)^{-1}$, 
$\tau_{\rm ex}$ is the time of exciton flipping between the bright states, 
$\tau_{\rm e}$ and $\tau_{\rm h}$ are the electron and hole spin flip times, 
and $\Delta$ is the splitting between the bright and dark exciton states. 
When $\Delta$ is smaller than $k_{\rm B}T$, the rate equations describing these processes yield the following expression for the relaxation time $\tau_P$ of the PL polarization: $\tau_P^{-1} = 2(\tau_{\rm e} + \tau_{\rm h})^{-1} + \tau_{\rm ex}^{-1}$~\cite{Leonard2009}. $\Delta$ is typically significantly smaller than $k_{\rm B}T$ for IXs~\cite{Leonard2009}. 
$\tau_{\rm h}$ is typically significantly smaller than $\tau_{\rm e}$ for GaAs heterostructures~\cite{Maialle1993}, whereas $\tau_{\rm e}$ is typically significantly smaller than $\tau_{\rm h}$ for MoSe$_2$/WSe$_2$ heterostructures~\cite{Zhou2017}, which aligns with the similarity between the schematics in Figs.~S10a and S10b.

\begin{figure}
\begin{center}
\includegraphics[width=8.5cm]{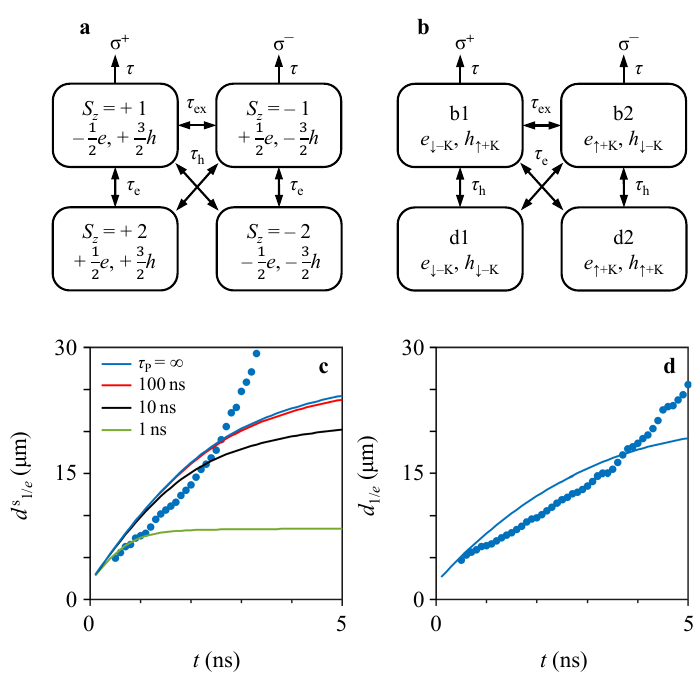}
\caption{Spin relaxation. (a,b) Schematic of spin relaxation processes in GaAs heterostructures (a) and H-stacked MoSe$_2$/WSe$_2$ heterostructures (b). (c) The measured spin transport kinetics (blue points) and spin transport kinetics simulated with Eq.~S5 for different polarization relaxation times $\tau_P$ (lines). (d) The measured IX transport kinetics (blue points) and IX transport kinetics simulated with Eq.~S4 (line). In the simulations, the mobilities for spin transport for infinite $\tau_P$ and for IX transport are estimated by a least-squares fit of the expansion kinetics simulated with Eqs.~S5 and S4 to the measured $d^{\rm s}_{1/e}(t)$ and $d_{1/e}(t)$, respectively.
$P_{\rm ex} = 500$~$\mu$W. $T = 3.5$~K.
}
\end{center}
\label{fig:spectra}
\end{figure}

Combining the spin relaxation equation (Eq.~S1) with the transport equation (Eq.~1) yields four coupled equations
\begin{eqnarray}
\frac{\partial {\bar n}}{\partial t} = \nabla \left[D \nabla {\bar n} + \mu {\bar n} \nabla (n_{\rm t} u_0 + U) \right] +
W {\bar n} + \bar{\Lambda}\
\end{eqnarray}
where $n_{\rm t} = n_{\rm d1} + n_{\rm b1} + n_{\rm b2} + n_{\rm d2}$. This set of equations can be simplified for the case of fast electron spin relaxation 
($\tau_{\rm e} \ll \tau_{\rm h}, \tau_{\rm ex}$; 
$\nabla \left[D \nabla {\bar n} + \mu {\bar n} \nabla (n_{\rm t} u_0 + U) \right] \ll W_{\rm e}{\bar n}$). 
This simplification leads to the following equations 
\begin{align}
\frac{\partial n_{\rm b}}{\partial t} &= 2 \nabla \left[D \nabla n_{\rm b} + \mu n_{\rm b} \nabla (2n_{\rm b} u_0 + U) \right] + \Lambda_{\rm b} - \frac{n_{\rm b}}{\tau} \\ 
\frac{\partial n_{\rm s}}{\partial t} &= 2 \nabla \left[D \nabla n_{\rm s} + \mu n_{\rm s} \nabla (2n_{\rm b} u_0 + U) \right] + \Lambda_{\rm s} - \frac{n_{\rm s}}{\tau} - \frac{n_{\rm s}}{\tau_P}\ 
\end{align}

Equation~S4 describes the transport of IXs with the density of bright IXs $n_{\rm b} = n_{\rm b1} + n_{\rm b2}$, which is characterized by the propagation of $I = I_{\sigma^+} + I_{\sigma^-}$. $\Lambda_{\rm b} = \Lambda^+ + \Lambda^-$. The factors of 2 appear due to dark excitons~\cite{Leonard2009}. 

Equation~S5 describes the transport of spin polarization density $n_{\rm s} = n_{\rm b1} - n_{\rm b2}$, which is characterized by the propagation of $I_{\rm spin} = I_{\sigma^+} - I_{\sigma^-}$. The last term $ - n_{\rm s} /\tau_P$ describes the polarization relaxation. $\Lambda_{\rm s} = \Lambda^+ - \Lambda^-$. 

Figure~S10c shows the measured spin transport kinetics (blue points) and the spin transport kinetics simulated with Eq.~S5 for different $\tau_P$ (lines). An increase of $\tau^{-1}_P$ suppresses spin transport (Fig.~S10c). In contrast, a comparison with the measured IX transport kinetics (blue points in Fig.~S10d) shows that IX-mediated spin transport is characterized by the spin mobility exceeding the IX mobility. This confirms that the effect of spin relaxation on the suppression of polarization transport is negligible during IX transport across the entire heterostructure ($\sim 10$~$\mu$m), which is consistent with the qualitative outline above.

\vskip3mm
$^*$ These authors contributed equally: W.~J.~Brunner, Zhiwen~Zhou.